\documentclass[english,aps,floats,twocolumn,showpacs,nofootinbib]{revtex4}
\usepackage{pslatex}
\usepackage[T1]{fontenc}
\usepackage[latin1]{inputenc}
\usepackage{graphicx}
\usepackage{epsfig}

\usepackage{calc}
\usepackage{ifthen}

{
{
{
\newcommand{\bea}{\begin{eqnarray}}
\newcommand{\eea}{\end{eqnarray}}

\newcommand{\nc}{\newcommand}
\nc{\renc}{\renewcommand}
\nc{\eqs}[2]{\mbox{Eqs.~(\ref{#1},\,\ref{#2})}}
\nc{\eq}[1]{\mbox{Eq.~(\ref{#1})}}
\nc{\figs}[2]{\mbox{Figs.~(\ref{#1},\,\ref{#2})}}
\nc{\fig}[1]{\mbox{Fig~.(\ref{#1})}}
\nc{\be}[1]{\begin{equation} \mbox{$\label{#1}$}}
\nc{\ee}{\vspace{0.1cm}\end{equation}}

\newcommand{\bean}{\begin{eqnarray*}}
\newcommand{\eean}{\end{eqnarray*}}

\def\GeV{{\rm \ GeV}}
\def\MeV{{\rm \ MeV}}

\def\bfx{{\bf x}}
\def\bfk{{\bf k}}

\def\tM{\tilde{M}}

\def\lae{\;^{<}_{\sim} \;} \def\gae{\; ^{>}_{\sim} \;}

\begin{document}
\title{Affleck-Dine Baryogenesis, Condensate Fragmentation and Gravitino Dark Matter in Gauge-Mediation with a Large Messenger Mass}
\author{Francesca Doddato}
\email{f.doddato@lancaster.ac.uk}
\author{John McDonald}
\email{j.mcdonald@lancaster.ac.uk}
\affiliation{Cosmology and Astroparticle Physics Group, Dept. of Physics, University of 
Lancaster, Lancaster LA1 4YB, UK}
\begin{abstract}

      We study the conditions for successful Affleck-Dine baryogenesis and the origin of gravitino dark matter in GMSB models. AD baryogenesis in GMSB models is ruled out by neutron star stability unless Q-balls are unstable and decay before nucleosynthesis. Unstable Q-balls can form if the messenger mass scale is larger than the flat-direction field $\Phi$ when the condensate fragments. We provide an example based on AD baryogenesis along a $d = 6$ flat direction for the case where $m_{3/2} \approx 2 \GeV$, as predicted by gravitino dark matter from Q-ball decay. Using a phenomenological GMSB potential which models the $\Phi$ dependence of the SUSY breaking terms, we numerically solve for the evolution of $\Phi$ and show that the messenger mass can be sufficiently close to the flat-direction field when the condensate fragments. We compute the corresponding reheating temperature and the baryonic charge of the condensate fragments and show that the charge is large enough to produce late-decaying Q-balls which can be the origin of gravitino dark matter.

\end{abstract}
\maketitle

\section{Introduction}

         Supersymmetry (SUSY) is widely viewed as the most likely theory beyond the Standard Model (SM). In SUSY extensions of the SM, the spectrum of sparticle masses is strongly dependent on the mechanism for SUSY breaking. Gauge- and gravity-mediated SUSY breaking are the most commonly considered mechanisms. In the case of gauge-mediated SUSY breaking (GMSB), the gravitino can be much lighter than the mass of the sparticles, making it a natural candidate for the lightest SUSY partner (LSP) and dark matter. For a complete cosmology of GMSB, we also need a mechanism for baryogenesis. In the MSSM, Affleck-Dine (AD) baryogenesis \cite{ad} is a particularly simple and effective way to generate the baryon asymmetry. A complex flat direction field gains a large expectation value, which later begins coherent oscillations in the real and imaginary directions. The interaction of the evolving field with B violating operators induces a net asymmetry in the flat-direction condensate. 

     It is well-known that AD condensates are unstable with respect to spatial perturbations \cite{ks,km1,km2,km3,kk1,kk2,kk3,fk}. These grow and fragment the condensate, producing Q-balls. The cosmology of Q-balls depends on the dimension of the B-violating operator and the nature of SUSY breaking. In gravity-mediated models the Q-balls are unstable and, if their charge is large, these may decay below the freeze-out temperature of neutralino dark matter \cite{km1,km2}. In this case dark matter and baryon number are simultaneously produced when the Q-balls decay, with $n_{DM} = 3 n_{B}$ by R-parity conservation. Therefore $\Omega_{B}/\Omega_{DM} = m_{n}/(3 m_{DM})$, where $m_{n}$ is the nucleon mass. For neutralino dark matter with mass $\gae 100 \GeV$, as expected in gravity-mediated SUSY breaking models, this produces a value much smaller than the observed baryon-to-dark matter ratio $\approx 1/6$, which requires $m_{DM} \approx 2 \GeV$. In this case the neutralino DM candidates must annihilate rapidly after production, pointing to Higgsino-like dark matter \cite{yamag}. 

          While this is an appealing mechanism for non-thermal dark matter, the possibility of directly producing both dark matter and the baryon asymmetry from Q-ball decay is lost. This is only possible with a dark matter candidate of mass $\approx$ 2 GeV. In \cite{rs} a light axino LSP in gravity-mediated SUSY breaking was proposed. Alternatively, in GMSB models a natural low-mass DM candidate is the gravitino. 

    However, Q-ball formation in GMSB AD baryogenesis is difficult to achieve without phenomenological problems \cite{kssw1,kssw2,shoeQ}. The flat-direction potential depends on the messenger mass. For field values much larger than the messenger mass, the potential becomes almost flat due to the suppression of gauge-mediation by the flat-direction field. If the magnitude of the flat-direction field when Q-balls form is much larger than the messenger mass, then the Q-balls will have an energy-per-charge, $E/Q$, which decreases as $Q^{-1/4}$ \cite{ks}. For large enough $Q$, $E/Q$ is less than the nucleon mass, implying that the Q-balls are absolutely stable if they carry baryon number. This leads to severe problems. A stable Q-ball captured by a neutron star will absorb baryon number, reducing its mass and eventually destabilizing the neutron star. This severely constrains AD baryogenesis in such GMSB models, ruling out all flat directions lifted by conventional B-conserving non-renormalizable potential terms \cite{kssw1}. More precisely, $E/Q \propto Q^{-1/4}$ is true if the AD field $\phi$ in the Q-ball is on the approximately constant part of the potential. (These are referred to as flat direction (FD) Q-balls in \cite{kssw1}.)  For sufficiently large Q-ball charge, $\phi$ becomes large enough for the non-renormalizable terms to become important in the Q-ball solution. This happens once $Q > Q_{c}$, where for the $d = 6$ Q-balls we consider $Q_{c} \sim 10^{24}$ \cite{kssw1}. For larger charge, the Q-balls become thin-walled, with constant field magnitude $\phi_{c} \sim 10^{14} \GeV$ inside and radius increasing as $Q$ increases. (These are referred to as curved direction (CD) Q-balls in \cite{kssw1}.) If Q-balls have a density comparable to dark matter (as we would expect if the Q-balls are stable and the baryon asymmetry is due to Affleck-Dine baryogenesis), then they will be excluded by neutron star stability if the non-renormalizable lifting terms conserve baryon number. FD type Q-balls imply a neutron star lifetime $\sim 10^{10}$ years, whereas CD Q-balls imply a lifetime $\sim 1500$ years \cite{kssw1}. The initial charge of Q-balls from AD baryogenesis will typically be less than $Q_{c}$ and therefore they will initially be FD type Q-balls. However, since $Q_{c} \ll Q_{ns}$, where $Q_{ns} \sim 10^{57}$ is the baryon number of a neutron star, the charge of any FD Q-ball trapped by a neutron star will rapidly grow by baryon absorption to become larger than $Q_{c}$ (for a constant absorption rate, this will occur in a time $\sim Q_{c}/Q_{ns} \times 10^{10}$ years $\sim$ $10^{-15}$ s), transforming them in CD Q-balls and destabilizing the neutron star.

         AD baryogenesis in GMSB could work if the messenger mass is sufficiently large that condensate fragmentation occurs when the amplitude of the flat-direction scalar is in the approximately quadratic part of the potential. In this case the resulting Q-balls will be unstable. It may then be possible to produce dark matter via late decay of Q-balls to out-of-equilibrium NLSPs which subsequently decay to gravitino with mass $m_{3/2} \approx 2 \GeV$.\footnote{An alternative mechanism for achieving unstable Q-balls in GMSB models is to produce Q-balls with a small baryonic charge, $10^{12} \lae Q \lae 10^{18}$. In this case the large suppression of the mass of the scalars forming the Q-ball does not occur and the Q-balls can decay directly to out-of-equilibrium gravitinos \cite{ksh1}.}

    In the case of gravity-mediated SUSY breaking, it is known that $d = 6$ flat directions lifted by superpotential terms of the form $(u^{c}d^{c}d^{c})^2$ can produce Q-balls with large enough baryon number to decay after neutralino freeze-out but before nucleosynthesis \cite{km1,km2}. We may expect similar behaviour in the case of GMSB flat directions with sufficiently large messenger mass. Therefore in the following we will focus on the example of $d=6$ flat directions 
\footnote{$d = 4$ directions are disfavoured for AD baryogenesis as the $d =4$ superpotential terms $QQQL$ and $u^{c}u^{c}d^{c}e^{c}$ cause too rapid proton decay, although this does not exclude AD leptogenesis along the $(H_{u}L)^2$ direction. 
$d =4$ flat directions may also thermalize before AD baryogenesis can occur \cite{cmr}.
$d =5$ directions require superpotential terms which violate R-parity, but LSP dark matter assumes R-parity conservation.}. Our method will be to numerically study a phenomenological GMSB flat-direction potential which models the suppression of the SUSY-breaking masses at field values greater than the messenger scale.

    Our paper is organized as follows. In Section 2 we introduce the phenomenological GMSB flat-direction potential. In Section 3 we consider AD baryogenesis and condensate fragmentation along a $d = 6$ MSSM flat-direction in a GMSB model with $m_{3/2} \approx 2 \GeV$. We solve the field equations to obtain the baryon density and so determine the reheat temperature. We then perform a semi-analytical study of condensate fragmentation using the phenomenological potential and estimate the time of formation and baryonic charge of the condensate fragments. In Section 4 we present our conclusions.

\section{Flat-direction potential in GMSB}

   GMSB models are based on SUSY breaking in a hidden sector which is transmitted to the MSSM via vector pairs of messenger fields with SM gauge charges. The messenger superfield scalar components acquire SUSY breaking mass splittings from their interaction with the hidden sector. The messengers then induce masses for MSSM gauginos at 1-loop and soft SUSY breaking scalar mass squared terms at 2-loops.   

     This is true so long as the flat-direction field $\Phi$ does not give the masses to the gauge fields which are greater than the messenger mass $M_{m}$. Once $|\Phi| \gae M_{m}$, the transmission of SUSY breaking via gauge fields is suppressed\footnote{More accurately, $g |\Phi| \gae M_{m}$, where $g$ is the relevant gauge coupling. For the $(u^{c}d^{c}d^{c})^2$ direction of interest to us here, $g$ is the strong coupling. Since $g \approx 1$, for simplicity we will not include $g$ explicitly in the potential.}. In general, all SUSY breaking mass terms and A-terms are expected to be proportional to the order parameter of SUSY breaking, $<F_{S}>$, therefore once $|\Phi| \gae M_{m}$ all SUSY breaking mass parameters are expected to be proportional to $<F_{S}>/|\Phi|$ \cite{deg}. Thus we must include a $|\Phi|^{-1}$ suppression factor for the soft SUSY breaking mass terms at large $|\Phi|$. A 2-loop calculation shows that the flat-direction effective potential has in fact a squared logarithmic dependence on $|\Phi|$ \cite{deg}.        
 
   Therefore to model $d = 6$ AD baryogenesis in GMSB models, we consider the following phenomenological flat-direction potential 
$$ V(\Phi) = m_{s}^2 M_{m}^2 \ln^{2} \left(1 + \frac{|\Phi|}{M_{m}}\right)\left(1 + K \ln \left( \frac{|\Phi|^2}{M_{m}^2} \right) \right) $$
 $$+ m_{3/2}^2\left(1 + \hat{K} \ln \left( \frac{|\Phi|^2}{M_{m}^2} \right) \right)|\Phi|^2 
   - cH^2 |\Phi|^2 + $$
  \be{e1}
(A W + h.c.) + 
\left| \frac{\Phi^{5}}{5! \tM^{3}}\right|^2         ~,\ee 
where
\be{e2}   W = \frac{\Phi^6}{6! \tM^3}     ~.\ee 
We include the factor $6!$ in \eq{e2} so that the physical strength of the interactions is dimensionally of the order of $\tM$. The first term in \eq{e1} is due to GMSB with messenger mass $M_{m}$. The factor multiplying this takes into account 1-loop radiative corrections due to gaugino loops once $|\Phi| \lae M_{m}$, with $K \approx -(0.01-0.1)$ \cite{km1,km2,km3}. The second term is due to gravity-medated SUSY breaking including the 1-loop correction term $ \hat{K}$. (For simplicity we set $\hat{K} = K$.) We have also included a Hubble correction to the mass squared term. We do not include a Hubble correction to the A-term, as the interaction leading to such terms is typically proportional to the inflaton field and so averages to zero over its coherent oscillations \cite{kawaA}.  The scale of the non-renormalizable terms, $\tM$, will depend on the origin of the superpotential term responsible for lifting the flat direction. This could be gravitational, $\tM \sim M_{p}$ (where $M_{p} = 2.4 \times 10^{18} \GeV$) or due to the new physics suggested by MSSM coupling constant unification, $\tM \sim M_{U} \approx 10^{16} \GeV$. In the following we will focus on the case where $\tM \sim M_{p}$.
For the A-term we consider
\be{e3}    A = m_{3/2} +  \frac{a_{o} m_{s}}{\left(1 + 
\frac{|\Phi|^2}{M_{m}^{2}}\right)^{1/2}  }          ~.\ee
The first term in \eq{e3} represents the A-term due to gravity-mediated SUSY breaking. The second term models the A-term in gauge-mediated SUSY breaking at $|\Phi| \lae M_{m}$, which is generated at 1-loop from the gaugino masses and is therefore suppressed relative to the A-term in gravity-mediated models, with $a_{o} \sim 0.01$. The suppression factor $(1 + |\Phi|^2/M_{m}^{2})^{-1/2}$ models the $1/|\Phi|$ suppression of the GMSB A-term at $|\Phi| \gg M_{m}$. 

  As a reference model for GMSB we will consider a metastable
SUSY breaking sector \cite{mura}, although our results depend only on the SUSY breaking 
contribution to the messenger scalar masses and therefore can easily be adapted to GMSB models more generally. 
The superpotential of the SUSY breaking sector is 
\be{e4}  W_{SB} = - \mu^2 S + \kappa S f \overline{f} + M f \overline{f}   ~,\ee
where $S$ is the SUSY breaking field and $f$, $\overline{f}$ are the messenger fields.
The K\"ahler potential is 
\be{e5}  K = |S|^2 - \frac{|S|^4}{4 \Lambda^2} + O\left( \frac{|S|^6}{\Lambda^4} \right)    ~,\ee
which is the generic form of K\"ahler potential for GMSB in a metastable vacuum.  
The potential is then 
$$  V = |\mu^2 - \kappa f \overline{f} |^2 \left(1 + \frac{|S|^2}{\Lambda^2}  
+ O\left(\frac{|S|^4}{\Lambda^4}\right) \right)  $$
\be{e6} + | \kappa S f + M f|^2  +  | \kappa S \overline{f} + M \overline{f}|^2 
~.\ee    
There is a local SUSY breaking minimum at $S = f = \overline{f} = 0$ if $M^2 > \kappa \mu^2$, in which case $F_{S} = \mu^2$. At the local SUSY breaking minimum the SUSY messenger mass and SUSY breaking mass squared splitting are 
\be{e7} M_{m} = M  \;\;\;\;\; ; \;\;\;\; F_{m} = \kappa F_{S} = \kappa \mu^2   ~.\ee
The masses squared of the scalar components of the messengers $f$, $\overline{f}$ are therefore $m^2_{f} = m^2_{\overline{f}} = M_{m}^2 \pm \kappa \mu^2$, while the fermion mass is $M_{m}$. Soft SUSY-breaking masses for gauginos and scalars in the MSSM sector of order $m_{s}$ are then generated by messenger loops 
\be{e8}  m_{s} \approx \frac{g^2}{16 \pi^2} \frac{F_{m}}{M_{m}} \equiv \frac{g^2}{16 \pi^2} \frac{\kappa \mu^2}{M_{m}}    ~.\ee
The gravitino mass is 
\be{e9} m_{3/2} = \frac{F_{S}}{\sqrt{3} M_{p}}  =  \frac{\mu^2}{\sqrt{3} M_{p}}       ~.\ee
Therefore the messenger mass is related to $m_{s}$ and $m_{3/2}$ by  
\be{e10} M_{m} \approx \frac{g^2}{16 \pi^2} \frac{\sqrt{3} \kappa m_{3/2} M_{p}}{m_{s}}    ~.\ee
Thus 
\be{e11}  M_{m} \approx 5 \times 10^{13} g^2 \kappa \left( \frac{m_{3/2}}{2 \GeV} \right) 
\left( \frac{100 \GeV}{m_{s}} \right)  \GeV      ~.\ee
From this we see that a large messenger mass scale is required to have $m_{3/2} \approx 2 \GeV$. The importance of this for $d = 6$ AD baryogenesis is that, for plausible values of $\tilde{M}$, the messenger mass can be comparable to the initial value of the flat direction scalar when 
$\Phi$ begins to oscillate and the baryon asymmetry is generated. 
To get a rough estimate of the value of $|\Phi|$ at the onset of oscillations, $|\Phi|_{osc}$, we compute the value of the minimum of $V(\Phi)$ for the case where only the $-c H^2 |\Phi|^2$ term and the SUSY non-renormalizable term are included. Setting $cH^2 = m_{s}^2$, the value when the potential is destablized by the mass squared term, then gives 
$$ |\Phi|_{osc} \sim \left(\frac{5!^{2}}{5} \right)^{1/8} \left(m_{s}^2 \tM^{6} \right)^{1/8}$$  
\be{e12} = 3 \times 10^{14} \GeV \times 
 \left( \frac{m_{s}}{100 \GeV} \right)^{1/4} \left( \frac{\tM}{10^{18} \GeV} \right)^{3/4} ~.\ee 
The value of $\tM$ is a free parameter in AD baryogenesis models. A natural possibility is that the non-renormalizable terms are associated with the completion of the theory at the Planck scale. However, even with this assumption there are different possibilities for how the Planck scale enters. If one expects the strength of the interaction to be dimensionally determined by $M_{P}$, then the factorial term should be included in the superpotential and so $\tM \approx M_{P}$. But it is also often assumed that the Planck scale enters in the Lagrangian rather than the vertex, in which case $W = \Phi^6/M_{p}^3$ and so  $\tM \approx M_{p}/(6!)^{1/3} = 2.7 \times 10^{17} \GeV$. We will therefore consider a range of $\tM$ from $
2 \times 10^{17}\GeV$ to $M_{p}$\footnote{Such scales may also arise in the case where $\tM$ is due to exchange of heavy particles of mass $M_{U} \sim 10^{16} \GeV$, once suppression of the non-renormalizable operators due to couplings is included.}.  

  For the case where $g^2 \kappa \approx 1$, $m_{s} \approx 100 \GeV$ and $\tM = 2 \times 10^{17} \GeV$, the messenger mass scale is $\approx 5 \times 10^{13} \GeV$ while $|\Phi|_{osc} \approx 8 \times 10^{13} \GeV$. Therefore it is possible that the messenger mass scale can be close to the value of $|\Phi|$ at the onset of oscillations. In addition, there will be a significant time delay between the onset of oscillations and the eventual fragmentation of the AD condensate and Q-ball formation. Therefore even if $|\Phi|$ is greater than $M_{m}$ when oscillations begin, it is possible that $|\Phi|$ will be less than $M_{m}$ when the condensate fragments. If Q-balls form when $|\Phi| \lae M_m$, then $E/Q$ will not be strongly suppressed, in contrast to the case of conventional GMSB Q-balls with $|\Phi| \gg M_{m}$.

    Thus $d = 6$ flat directions, combined with $m_{3/2} \approx 2 \GeV$ and $\tM \approx 10^{17}-10^{18} \GeV$, may lead to condensate fragmentation in a region of the potential where unstable Q-balls can form. In order to better understand the evolution of the flat-direction condensate, in the next section we will numerically evolve $\Phi$ using the phenomenological GMSB potential given in \eq{e1}.

\section{Affleck-Dine Baryogenesis and Condensate Fragmentation}

\subsection{$d = 6$ AD Baryogenesis in GMSB with a large messenger mass}

    The baryon asymmetry in the late-time coherent oscillations of the real and imaginary parts of $\Phi$ is induced by the B-violating A-term. The baryon number density is $n_{B} = n_{Q}/3$, where $n_{Q}$ is the global $U(1)$ charge density corresponding to the number asymmetry in $\Phi$ particles:
\be{e12a} n_{Q} = i\left( \dot{\Phi}^{\dagger} \Phi - \Phi^{\dagger} \dot{\Phi} \right) \equiv \phi_{1} \dot{\phi}_{2}  -  \dot{\phi}_{1} \phi_{2}  ~.\ee
The factor $1/3$ accounts for the baryon number of $\Phi$ along the $(u^{c}d^{c}d^{c})^2$ direction.  
In the following we will use subscript $b$ to denote quantities at the time when the baryon asymmetry in a comoving volume becomes fixed. The present baryon asymmetry to entropy is then 
\be{e12b} \frac{n_{B}}{s} = \frac{n_{B\;b}T_{R}}{4 H_{b}^{2} M_{p}^2}  
~.\ee
The observed baryon asymmetry is $(n_{B}/s)_{obs} = 1.8 \pm 0.1 \times 10^{-10}$. This allows us to fix the reheating temperature for a given 
flat-direction potential by computing $H_{b}$ and $n_{B\;b}$.

     In Fig.1 we show the growth of the baryon asymmetry per comoving volume as a function of $|\Phi|/M_{m}$ for the case $M_{m} = 5 \times 10^{13} \GeV$, $m_{s} = 100 \GeV$, $K = -0.1$ and $\tM = M_{p}$, where $|\Phi|$ is the maximum value of the magnitude of $\Phi$ on the ellipitical trajectory. (We assume $a_{o} = 0.01$ in all our calculations.) This shows the rapid increase in the asymmetry once the $\Phi$ oscillations begin, with the asymmetry becoming fixed once $|\Phi|/M_{m} \lae 3$. In Fig.2 we show the trajectory of $\Phi$ in the complex plane when $|\Phi|/M_{m} = 0.54$. A noteworthy feature of this is that the trajectory is not yet ellipitical but in fact precesses, even though the baryon asymmetry per comoving volume is constant. This indicates that the GMSB deviation from a $|\Phi|^2$ potential is still significant at this time, even though the A-terms have become negligible and the baryon number in a comoving volume is constant.

\begin{figure}[htbp]
\begin{center}
\epsfig{file=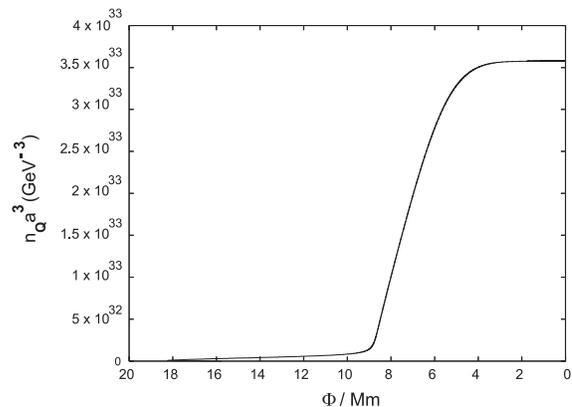, width=0.3\textwidth, angle = -90}
\caption{Growth of the baryon asymmetry per comoving volume as a function of $|\Phi|/M_{m}$.}
\label{fig1}
\end{center}
\end{figure}

\begin{figure}[htbp]
\begin{center}
\epsfig{file=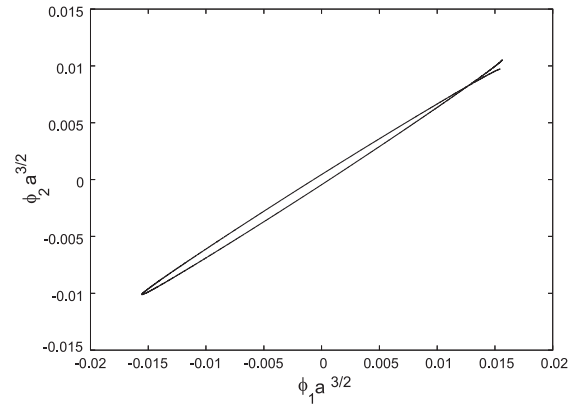, width=0.3\textwidth, angle = -90}
\caption{Trajectory of $\Phi$ in the complex plane after formation of the baryon asymmetry. The ellipse precesses but the baryon asymmetry per comoving voume is constant.}
\label{fig2}
\end{center}
\end{figure}

In Table 1 we give the AD baryogenesis parameters and $T_{R}$ for a range of $\tM$  and $M_{m}$ for the case $m_{s} = 100 \GeV$. In Table 2  we show the case with $m_{s} = 500 \GeV$. From this we see that in most cases the baryon asymmetry forms when $|\Phi|/M_{m}$ is larger than or of the order of 1.

\begin{table}[h]
\begin{center}
\begin{tabular}{|c|c|c|c|c|c|c|c|c|}

\hline $m_{s}$ &  $\tM $ & $M_{m} $ & & $n_{B\;b}$ & $H_{b}$ & 
$\frac{|\Phi|}{M_{m}}$ & $T_{R}$  \\ 
	\hline   100 & $2.4 \times 10^{18}$  & $5 \times 10^{13}$ & & $1.4 \times 10^{28}$ & 6.8 & 2.5 & 14.4 \\
	\hline   100 & $2.4 \times 10^{18}$  & $5 \times 10^{12}$ & & $4.3 \times 10^{27}$ & 1.6 & 22.6 & 2.4 \\
	\hline   100 & $2.4 \times 10^{18}$  & $1 \times 10^{12}$ & & $4.3 \times 10^{25}$ & 0.4 & 75.0 & 14.7 \\
	\hline   100 & $2.0 \times 10^{17}$  & $5 \times 10^{13}$ & & $3.7 \times 10^{26}$ & 15.4 & 0.5 & 2701 \\
	\hline   100 & $2.0 \times 10^{17}$  & $5 \times 10^{12}$ & & $2.9 \times 10^{26}$ & 5.7 & 14.1 & 464 \\
	\hline   100 & $2.0 \times 10^{17}$  & $1 \times 10^{12}$ & & $1.2 \times 10^{26}$ & 2.0 & 17.3 & 140 \\

	\hline     
 \end{tabular}
 \caption{\footnotesize{AD baryogenesis parameters for $m_{s} = 100 \GeV$. (Dimensionful quantities in GeV units.) }}  
 \end{center}
 \end{table}

\begin{table}[h]
\begin{center}
\begin{tabular}{|c|c|c|c|c|c|c|c|}

\hline $m_{s} $ &  $\tM $ & $M_{m} $ & $n_{B\;b} $ & $H_{b}$ & 
$\frac{|\Phi|}{M_{m}}$ & $T_{R}$  \\ 
	\hline   500 & $2.4 \times 10^{18}$  & $1 \times 10^{13}$ & $1.2 \times 10^{28}$ & 9.7 & 14.9 & 33.3 \\
	\hline   500 & $2.4 \times 10^{18}$  & $1 \times 10^{12}$ & $6.7 \times 10^{27}$ & 1.9 & 119.6 & 2.1 \\
	\hline   500 & $2.0 \times 10^{17}$  & $1 \times 10^{13}$ & $9.3 \times 10^{26}$ & 33.0 & 3.2 & $5.1 \times 10^{3}$ \\
	\hline   500 & $2.0 \times 10^{17}$  & $1 \times 10^{12}$ & $1.2 \times 10^{26}$ & 6.8 & 19.7 & 845 \\
	\hline     
 \end{tabular}
 \caption{\footnotesize{AD baryogenesis parameters for $m_{s} = 500 \GeV$. (Dimensionful quantities in GeV units.)  }}  
 \end{center}
 \end{table}

\subsection{Perturbation growth and condensate fragmentation}

   To study the growth of perturbations of the AD condensate we will use the approach of \cite{ks}. This assumes that the baryon asymmetry in the condensate is maximal i.e. the trajectory of $\Phi$ is circular in the absence of expansion ("Q-matter"). This is not true for GMSB models with $a_{o} \ll 1$ and $m_{3/2} \ll m_{s}$, which produce ellipitical trajectories. However, the growth rate of energy density perturbations for the ellipitical trajectory will be similar, since in this case the growth of perturbations will be determined by the negative pressure of the condensate averaged over oscillation cycles, which turns out to be essentially independent of the field trajectory for a given potential and maximum value of the field on the trajectory, $\phi_{max}$. The average pressure for the limiting case of a linear oscillating condensate (i.e. a real field) has been discussed in \cite{turner}, where the average pressure was shown to be 
\be{rx1}  \frac{<p>}{\rho}  + 1  = 2 \frac{\int_{0}^{\phi_{max}} \left( 1 - \frac{V}{V_{max}} \right)^{1/2} d \phi }{\int_{0}^{\phi_{max}} \left( 1 - \frac{V}{V_{max}} \right)^{-1/2} d \phi} ~ ~.\ee
Perturbations of the energy density of wavenumber ${\bf k}$ then grow according to 
\be{rx2}   \delta \ddot{\rho}_{{\bf k}} = -<p> |{\bf k}|^2 \delta \rho_{{\bf k}}   ~.\ee
In this we have neglected expansion since the growth rate is typically large compared with $H$.  
In the case of the circular condensate, the pressure is constant and determined by the magnitude of the field and the potential,
\be{rx3} \frac{p}{\rho} = \frac{\frac{\phi V^{'}}{2} - V}{\frac{\phi V^{'}}{2} + V}    ~.\ee
This expression allows us to derive 
simple expressions for the pressure and growth rate of perturbations which depend only on the potential and its derivatives, unlike the linear oscillation case where 
the average pressure must be calculated numerically via \eq{rx1}. 
The key observation that allows us to also use \eq{rx3} for non-circular trajectories is that the pressures in a circular condensate and in a linear condensate for a given $\phi_{max}$ are essentially identical. For the case of a polynomial potential, the pressures can be calculated exactly in both cases and are exactly equal,  $<p>/\rho = (n-2)/(n+2)$  for $V(\Phi) \propto |\Phi|^{n}$ \cite{turner}. For a non-polynomial potential there is no analytical 
expression and the integrals in \eq{rx1} must be calculated numerically. (This is not straightforward as the integrand in the denominator of \eq{rx1} is divergent as 
$V \rightarrow V_{max}$.) We have numerically integrated \eq{rx1} for the potential \eq{e1} and find that the negative pressure is equal to within 10$\%$ for a given $\phi_{max}$. 
Therefore the growth rate of perturbations of the circular and linear condensate will be essentially the same. We can expect the same to be true for ellipitical trajectories which are between these limiting cases. This allows us to use the more convenient expression for the pressure \eq{rx3} to calculate the growth of perturbations of the ellipitical condensate and so the time when they become non-linear. The similarity of the growth of perturbations in the circular and linear condensates is physically reasonable, as the growth of perturbations is essentially due to an attractive interaction between the scalar particles, and the strength of the interaction is the same for real and complex scalars which have the same potential.  This approach is also in agreement with the numerical results of \cite{kk3}, where it was found that the size of the condensate fragments in the circular and elliptical cases differs by a factor of at most 2.

      We first consider condensate fragmentation for a circular condensate in the potential \eq{e1}. 
The flat-direction field can be written as 
\be{e13} \Phi = \frac{1}{\sqrt{2}} R(\bfx,t) e^{i \Omega(\bfx,t)}   ~.\ee   
The $\Phi$ field equations are then 
\be{e14} \ddot{\Omega} + 3 H \dot{\Omega} - \frac{1}{a^2} \nabla^{2} \Omega + 2 \frac{\dot{R}}{R} \dot{\Omega} - \frac{2}{a^2 R} \nabla \Omega . \nabla R = 0       ~\ee
\be{e15} \ddot{R} + 3 H \dot{R} - \frac{1}{a^2} \nabla^2 R - \dot{\Omega}^2 R + \frac{1}{a^2} (\nabla \Omega)^2 R + V^{'} = 0    ~,\ee
where $V^{'} = dV/dR$. 
In this it has been assumed that $V$ is a function of $R$ alone, so these equations can be used to study growth of perturbations only once the A-terms are no longer significant in $V$ i.e. once $H < H_{b}$. The growing mode solutions have the form 
\be{e15a}  \delta R \;\;,\; \delta \Omega \propto e^{S(t) - i \bfk.\bfx}    ~.\ee
Substituting these into \eq{e13} and \eq{e14} and assuming that $\alpha = \dot{S} = constant$ gives \cite{ks}
$$ \left[ \alpha^2 + 3 H \alpha + \frac{\bfk^2}{a^2} + \frac{2 \dot{R}}{R} \alpha \right] \left[\alpha^2 + 3 H \alpha + \right.$$
\be{e16} \left. \frac{|\bfk|^2}{a^2}
-\dot{\Omega}^2 + U^{''}(R) \right] + 4 \dot{\Omega}^2 \left[ \alpha - \frac{\dot{R}}{R} \right] \alpha = 0    ~.\ee 
$\alpha$ is assumed to be constant because the rate of growth of perturbations is large compared with the rate of expansion of the Universe, in which case the time dependence in \eq{e16}, which is entirely due to expansion in the case of circular trajectory, can be neglected. In this case $\dot{R}$ can also be set to zero, as this is non-zero only because of expansion. In this case the equation simplifies to
\be{e16a} \left[ \alpha^2 + \frac{|\bfk|^2}{a^2} \right] \left[\alpha^2 + \frac{|\bfk|^2}{a^2}
-\dot{\Omega}^2 + V^{''}(R) \right] + 4 \dot{\Omega}^2 \alpha^2 = 0    ~.\ee 
This can be solved for $\alpha^2$
\be{e17}  \alpha^2 = \frac{|\bfk|^2}{a^2} \frac{1}{ \left( 
V^{''} + 3 \dot{\Omega}^2 \right) } \left( \dot{\Omega}^2 - 
V^{''} - 16 \frac{|\bfk|^2}{a^2} \frac{ \dot{\Omega}^4 } {\left(V^{''} + 3 \dot{\Omega}^{2}\right)^{2} } \right)     ~.\ee 
From this the largest value of $|\bfk|$ for which growth is possible is 
\be{e18} \left| \frac{\bfk_{max}}{a} \right|^2 \equiv K_{m}^2 \approx \left( \frac{4 \dot{\Omega}^2}{V^{''} + 3 \dot{\Omega}^2}\right) \times
\left(\dot{\Omega}^2 - V^{''}\right)   ~,\ee
where we have defined $K_{m} = \left|\bfk_{max}/a\right|$. 
For a homogeneous Q-matter background with $\dot{R} = \ddot{R} = 0$, \eq{e15} implies that $\dot{\Omega}^{2} = V^{'}/R$.  

      The solution for $\alpha$ in \eq{e17} gives the value of $\alpha(t)$ on a time-scale short compared to $H^{-1}$. 
One can then re-introduce the $t$ dependence of $R$ and $\dot{\Omega}$ due to the expansion of the Universe. The growth factor $S(t)$ is then given by 
\be{e19} S(t) = \int_{t_{*}}^{t} \alpha(t) dt   ~,\ee
where $t_{*}$ is the time at which a mode of wavenumber $\bfk$ begin to grow.

   For a given mode $\bfk$, the perturbation of the condensate will only begin to grow once $|\bfk/a| < K_m$. Suppose for a given mode this occurs at a time $t_{*}$. The subsequent growth of the perturbation follows from 
\be{e20}  S(t) = \int_{\alpha_{*}}^{\alpha}  \frac{\alpha(t)}{a H} da    ~,\ee   
where 
\be{e21} \alpha(\bfk,a)  
\approx \left( \frac{|\bfk|^2}{a^2} \frac{ \left( \dot{\Omega}^2 - V^{''} \right) }{ \left( V^{''} + 3 \dot{\Omega}^2 \right)}    \right)^{1/2}   ~.\ee
With $H \propto a^{-3/2}$ during inflaton-domination, we find from \eq{e19}
\be{e22} S(\bfk,a) = 2 \alpha(\bfk,a)  \left( 1 - \frac{a_{*}^{1/2}}{a^{1/2}} \right) H^{-1} ~.\ee 
 At a given value of $a$, the mode with the maximum growth is found by maximizing $S(\bfk,a)$ with respect to $|\bfk|$. Since $\alpha \propto |\bfk|$ and $ a_{*}(k) = K_{m}^{-1} |\bfk|$ 
we find
\be{e23} S(\bfk) \propto  |\bfk| - \frac{|\bfk|^{3/2}}{a^{1/2} K_{m}^{1/2}}    ~.\ee
This is maximized at 
\be{e24} \frac{|\bfk|_{frag}}{a} = \frac{4}{9} K_{m}  ~.\ee 

  Therefore at each value of $a$, the mode which has the largest growth will have wavenumber given by 
\eq{e24}. The value of $S$ for this mode is 
\be{e24a} S = \frac{2}{3} \alpha H^{-1}     ~.\ee
The condition for the condensate to fragment is  
\be{e25} \frac{\delta R}{R}  = \frac{\delta R_{o}}{R_{o}} e^{S} \; \gae 1 \; ~,\ee
for the mode with wavenumber \eq{e24}. The diameter of the fragments is then given by
\be{e25}   \lambda_{frag} \approx \frac{2 \pi }{\left(\left|\bfk\right|_{frag}/a\right)}    ~,\ee
while their global $U(1)$ charge ($Q = 3B$) is
\be{e25a}  Q \approx  n_{B} \frac{4 \pi}{3} \left( \frac{\lambda_{frag}}{2} \right)^{3}  ~.\ee

   We expect the same value of $S$ to apply to the elliptical trajectory when $R = R_{max}$, the maximum value of $R$ for the ellipse. We will therefore use this to estimate the time of fragmentation and the value of $|\Phi|/M_{m}$ at fragmentation. Given the energy density and baryon density when the condensate fragments, we can then compute the energy and baryon number of the fragments. 

 To compute the time of fragmentation of the condensate, we need a spectrum of initial perturbations. These are related to the primordial density perturbations. 
For the initial perturbation entering the horizon we expect $\delta R/R \approx \delta \Omega/\Omega \approx \delta \rho/\rho$, the usual primordial density perturbation. This is certainly true of the amplitude, since $R \propto H^{1/4}$ at the minimum of the potential
due to the Hubble mass correction. For the phase, once the perturbation enters the horizon we can expect field dynamics to 
transfer the perturbation of $R$ into a correlated $\Omega$ perturbation of a similar magnitude. There could also be additional phase fluctuations independent of the amplitude which will result in isocurvature baryon perturbations, but these are model dependent and we will not consider them here. 
With $\delta R_{o}/R_{o} \approx \delta \rho/\rho \approx 10^{-4}$, the condition for fragmentation becomes $S \gae 11$.  

   In Figure 3 we show the growth factor $S$ as a function of $|\Phi|/M_{m}$ for the case $m_{s} = 100 \GeV$, $K = -0.1$, $M_{m} = 5 \times 10^{13} \GeV$ and $\tM = M_{p}$. It can be seen that fragmentation occurs at $|\Phi|/M_{m} \approx 0.1$, even though the baryon asymmetry forms at a much larger value, $|\Phi|/M_{m} \approx 2.5$. 
In Table 3 we show the condensate fragmentation parameters for the case $m_{s} = 100 \GeV$ and in Table 4 for the case where $m_{s} = 500 \GeV$. We see that fragmentation can easily occur with $|\Phi|/M_{m} \lae 1$, in particular if $\tM < M_{p}$ and $|K| < 0.1$.  

    The field trajectory at fragmentation is strongly ellipitical, indicating that the total energy density in the $\Phi$ field is large compared with the energy density in Q-matter of the same charge density. We can estimate this ratio for a strongly elliptical trajectory by 
\be{e12c}  r_{E} \equiv  \frac{\rho}{\left(E/Q\right) n_{Q}} \approx \frac{V(|\Phi|)}{m_{s} n_{Q}} ~.\ee
This is correct for the case where $|\Phi|$ for the Q-matter configuration is sufficiently less than $M_{m}$ that it can be considered to be described by $|\Phi|^2$ potential with $E/Q \approx m_{s}$. Otherwise $E/Q$ will be less than $m_{s}$, so that \eq{e12c} will be an underestimate of $r_{E}$. $r_{E}$ also gives the ratio of the energy of the condensate fragments to the energy of a Q-ball of the same charge.   For the cases being considered we find that $r_{E}$ is mostly in the range 10-60 for $m_{s} = 100 \GeV$ and 20-90 for $m_{s} = 500 \GeV$.

    We next estimate the decay temperature of the Q-balls under the assumption that the condensate fragments evolve into gravity mediated-type Q-balls of the same charge, corresponding to Q-balls in a potential $V = m_{s}^{2} |\Phi|^2 (1  + K \ln (|\Phi|^{2})$. 
The decay rate of a Q-ball has an upper bound given by \cite{Qdecay}
\be{e26} \Gamma_{d} \leq \frac{\omega^3 A}{192 \pi^2 Q}  ~,\ee
where the Q-ball solution is proportional to ${\rm exp}(i \omega t)$ and $A$ is the area of the Q-ball. The upper limit is expected to be saturated for flat-direction Q-balls. The decay temperature is then  
\be{e27} T_{d} = \left( \frac{\omega^3 R^2 M_{p}}{48 \pi k_{T} Q} \right)^{1/2}   ~,\ee
where $k_{T} = (\pi^2 g(T)/90)^{1/2} \approx 4.7$ using $g(T) \approx 200$ for the MSSM. 
Dimensionally we expect $\omega \sim m$ and $R \sim m^{-1}$, so we can parameterize $\omega^3 R^2$ as $\omega^3 R^2 = f_{Q} m$. In particular, for gravity-mediated Q-balls $\omega \approx m$ and $R^2 = 2/(|K|m^2)$, therefore $f_{Q} = 2/|K|$. 
This assumes $\Phi$ decay to scalar pairs is kinematically excluded, otherwise the decay rate can be enhanced by a factor $f_{s} \sim 10^{3}$ since the decay is not Pauli-blocked within the volume \cite{km2}. The decay temperature can then be written as
\be{e28} T_{d} = 60 \MeV f_{Q}^{1/2} \left( \frac{m}{100 \GeV} \right)^{1/2} \left( \frac{10^{20}}{Q} \right)^{1/2}   ~.\ee

  In Table 5 we show the gravity-mediated type Q-ball properties for the case $m_{s} = 100 \GeV$ and in Table 6 for the case $m_{s} = 500 \GeV$.  In most cases the Q-ball decay temperature is well above the temperature of the onset of nucleosynthesis $\approx 1 \MeV$ but well below the typical temperature of NLSP freeze-out $\approx M_{LSP}/20 \gae 5 \GeV$.
(The decay temperature will be somewhat higher if $f_{s} > 1$.) Therefore such Q-balls can be a source of non-thermal NLSPs which subsequently decay to gravitino dark matter. 

    The condensate lumps will evolve into gravity-mediated Q-balls if at the end of their evolution $|\Phi|/M_{m} \ll 1$, 
so that the dynamics of the Q-ball are entirely due to the radiatively-corrected $|\Phi|^2$ potential. The initial 
values of $|\Phi|/M_{m}$ from Tables 3 and 4 are not always very much less than 1. However, since the energy in the initial condensate fragments is larger than the final energy in the Q-ball by a factor $r_{E}$, the field in the final Q-ball configuration 
will be smaller by a factor $\approx 1/\sqrt{r_{E}}$. In most cases this is enough to ensure $|\Phi|/M_{m} < 1$ for the Q-balls. 

   Note that if $|\Phi|/M_{m}$ is less than 1 but not very small, the Q-ball may correspond to a new type of Q-ball which is intermediate between the gravity- and gauge-mediated cases. We will study the structure of such intermediate Q-balls in a future discussion
\cite{dm2}.

\begin{table}[h]
\begin{center}
\begin{tabular}{|c|c|c|c|c|c|c|c|}
	\hline   $m_{s}$ &  $K$ &  $\tM $  & $M_{m} $ & $\frac{|\Phi|}{M_{m}}$
 & $K_{m}$ & $n_{B\;frag}$ & $r_{E}$  \\ 
      \hline   100 & -0.1 & $2.4 \times 10^{18}$ & $5 \times 10^{13}$ & 0.12  & 54.4 & $5.0 \times 10^{25}$   & 19.1 \\
      \hline   100 & -0.01 & $2.4 \times 10^{18}$ & $5 \times 10^{13}$ & 0.017  & 21.1 & $2.9 \times 10^{24}$   & 24.3  \\
      \hline   100 & -0.1 & $2.4 \times 10^{18}$ & $5 \times 10^{12}$ & 3.2 & 24.9 &   $3.0 \times 10^{26}$   & 5.8  \\
      \hline   100 & -0.1 & $2.4 \times 10^{18}$ & $1 \times 10^{12}$ & 12.3 & 9.6 &  $4.3 \times 10^{26}$    & 1.6  \\
	\hline   100 & -0.1 & $2.0 \times 10^{17}$ & $5 \times 10^{13}$ & 0.006  & 45.1 & $1.8 \times 10^{23}$   & 43.3  \\
	\hline   100 & -0.01 & $2.0 \times 10^{17}$ & $5 \times 10^{13}$ & 0.0009  & 14.6 &  $1.1 \times 10^{21}$   & 57.9  \\   
      \hline   100 & -0.1 & $2.0 \times 10^{17}$ & $5 \times 10^{12}$ & 0.26  & 55.1 &  $2.7 \times 10^{24}$   & 16.3  \\
      \hline   100 & -0.1 & $2.0 \times 10^{17}$ & $1 \times 10^{12}$ & 2.3  & 29.5 &  $7.0 \times 10^{24}$   & 7.3  \\

\hline     
 \end{tabular}
 \caption{\footnotesize{Condensate fragmentation parameters for $m_{s} = 100 \GeV$. (Dimensionful quantities in GeV units.)}}  
 \end{center}
 \end{table}

\begin{table}[h]
\begin{center}
\begin{tabular}{|c|c|c|c|c|c|c|c|}
\hline   $m_{s} $ &  $K$ &  $\tM$  & $M_{m}$ & $\frac{|\Phi|}{M_{m}}$
 & $K_{m}$ & $n_{B\;frag}$ & $r_{E}$  \\ 
       \hline  500 & -0.1 & $2.4 \times 10^{18}$ & $1 \times 10^{13}$ & 2.2 & 149.9 &  $7.3 \times 10^{26}$      & 31.8 \\
       \hline  500 & -0.01 & $2.4 \times 10^{18}$ & $1 \times 10^{13}$ & 2.2 & 153.3 & $7.3 \times 10^{26}$    & 31.8 \\
       \hline  500 & -0.1 & $2.4 \times 10^{18}$ & $1 \times 10^{12}$ & 21.2 & 30.6 & $8.3 \times 10^{26}$    & 1.9 \\
       \hline  500 & -0.1 & $2.0 \times 10^{17}$ & $1 \times 10^{13}$ & 0.15 & 276 & $4.7 \times 10^{24}$     & 72.2  \\    
       \hline  500 & -0.01 & $2.0 \times 10^{17}$ & $1 \times 10^{13}$ & 0.03 & 118 &   $1,2 \times 10^{23}$   & 85.3  \\
       \hline  500 & -0.1 & $2.0 \times 10^{17}$ & $1 \times 10^{12}$ & 3.8 & 111 & $2.1 \times 10^{25}$     & 20.0  \\

\hline     
 \end{tabular}
 \caption{\footnotesize{Condensate fragmentation parameters for $m_{s} = 500 \GeV$. (Dimensionful quantities in GeV units.)}}  
 \end{center}
 \end{table}

 \begin{figure}[htbp]
\begin{center}
\epsfig{file=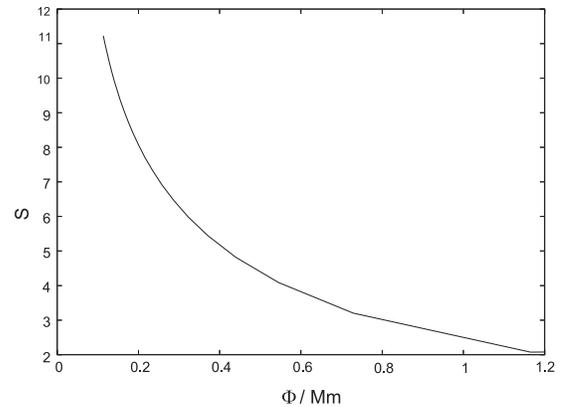, width=0.3\textwidth, angle = -90}
\caption{Growth factor S(\emph{t}) as a function of $|\Phi|/M_{m}$.}
\label{fig2}
\end{center}
\end{figure}


\begin{table}[h]
\begin{center}
\begin{tabular}{|c|c|c|c|c|c|c|}
	\hline   $m_{s}$ &  $K$ &  $\tM$  & $M_{m}$ & $Q$
 & $T_{d}(\MeV) $  \\ 
      \hline   100 & -0.1 & $2.4 \times 10^{18}$ & $5 \times 10^{13}$ & $1.1 \times 10^{23}$  & 8.1 \\
 \hline   100 & -0.01 & $2.4 \times 10^{18}$ & $5 \times 10^{13}$ & $4.0 \times 10^{22}$  & 42 \\
      \hline   100 & -0.1 & $2.4 \times 10^{18}$ & $5 \times 10^{12}$ & $8.0 \times 10^{24}$ & 1.0   \\
      \hline   100 & -0.1 & $2.4 \times 10^{18}$ & $1 \times 10^{12}$ & $6.5 \times 10^{25}$ & 0.3   \\
\hline   100 & -0.1 & $2.0 \times 10^{17}$ & $5 \times 10^{13}$ & $2.6 \times 10^{20}$  & 167 
\\
	\hline   100 & -0.01 & $2.0 \times 10^{17}$ & $5 \times 10^{13}$ & $1.3 \times 10^{20}$  & 236  \\   
      \hline   100 & -0.1 & $2.0 \times 10^{17}$ & $5 \times 10^{12}$ & $6.5 \times 10^{21}$  & 33    \\
      \hline   100 & -0.1 & $2.0 \times 10^{17}$ & $1 \times 10^{12}$ & $9.2 \times 10^{25}$  & 0.3   \\

\hline     
 \end{tabular}
 \caption{\footnotesize{Q-ball charge and decay temperature for $m_{s} = 100 \GeV$. (Dimensionful quantities in GeV units except $T_{d}$.)}}  
 \end{center}
 \end{table}

\begin{table}[h]
\begin{center}
\begin{tabular}{|c|c|c|c|c|c|c|}
	\hline   $m_{s}$ &  $K$ &  $\tM$  & $M_{m}$ & $Q$
 & $T_{d} (\MeV)$   \\ 
      \hline   500 & -0.1 & $2.4 \times 10^{18}$ & $1 \times 10^{13}$ & $7.8 \times 10^{22}$  & 22   \\
      \hline   500 & -0.01 & $2.4 \times 10^{18}$ & $1 \times 10^{13}$ & $7.8 \times 10^{22}$ & 69    \\
      \hline   500 & -0.1 & $2.4 \times 10^{18}$ & $1 \times 10^{12}$ & $1.1 \times 10^{25}$  & 1.8   \\
    \hline   500 & -0.1 & $2.0 \times 10^{17}$ & $1 \times 10^{13}$ & $7.7 \times 10^{19}$  & 683 \\
	\hline   500 & -0.01 & $2.0 \times 10^{17}$ & $1 \times 10^{13}$ & $2.7 \times 10^{19}$  & 3651  \\   
      \hline   500 & -0.1 & $2.0 \times 10^{17}$ & $1 \times 10^{12}$ & $5.7 \times 10^{21}$  & 79     \\

\hline     
 \end{tabular}
 \caption{\footnotesize{Q-ball charge and decay temperature for $m_{s} = 500 \GeV$. (Dimensionful quantities in GeV units except $T_{d}$.)}}  
 \end{center}
 \end{table}

\section{Conclusions}

    In this paper we have revisited Affleck-Dine baryogenesis and Q-ball formation in GMSB models. We have been concerned with two questions: (i) whether unstable Q-balls can naturally form in GMSB AD baryogenesis models and (ii) whether such unstable Q-balls can be a source of gravitino dark matter. To answer these questions we have focused on a $d = 6$ flat direction with $m_{3/2} = 2 \GeV$, which is consistent with gravitino dark matter from Q-ball decay.

    We find that it is possible for condensate fragmentation in $d = 6$ AD baryogenesis to occur when $|\Phi|/M_{m}$ is less than or of the order of 1. The importance of this is that fragmentation does not occur on the plateau of the GMSB  potential and so does not lead to formation of stable Q-balls which would destabilize neutron stars. Instead the fragments form either in the intermediate region between the 
$|\Phi|^2$ region and the GMSB plateau or well within the $|\Phi|^2$ dominated region, where we can expect the Q-balls to behave like unstable gravity-mediated Q-balls. If the condensate fragments subsequently evolve into Q-balls of the same charge, then the Q-balls should be of approximately gravity-mediated form and unstable, with decay temperatures typically in the range 10 MeV - 1 GeV. Decay of such Q-balls to NLSPs could then be a non-thermal source of gravitino dark matter. Even if $|\Phi|/M_{m} \sim 1$  when the condensate fragments, the final value of $|\Phi|/M_{m}$ in the Q-ball will typically be significantly less than 1 once the fragment has lost its excess energy, since the energy of the initial condensate fragment is much larger than the energy in its baryonic charge. 

     In some cases the value of $|\Phi|$ will be less than the messenger scale but not sufficiently small that the Q-ball will be a gravity-mediated type Q-ball. In this case we need to consider a new type of Q-ball, intermediate between gauge- and gravity-mediated Q-balls, whose structure will be determined by the region of the potential where the 
corrections due to the messenger scale are still significant.  We will study the structure of these intermediate Q-balls in a future analysis \cite{dm2}. 

    Due to the suppressed A-terms in GMSB models, the energy in the condensate fragments is typically larger than the energy in the baryon asymmetry by a factor of 10-100. 
In our discussion we have assumed that the condensate fragments of charge Q subsequently evolve into Q-balls of the same charge. 
One may question whether this assumption is reasonable. We have considered condensate fragments which initially form with only a small amount of energy in their baryonic charge relative to their mass. This is consistent with the most recent numerical simulation of condensate fragmentation, presented in \cite{fk}, where the initial condensate fragments are called "first generation Q-balls". Their subsequent evolution is found to depend on how ellipitical the condensate trajectory is. It was found that if $r_{E}$ (essentially equivalent to $\epsilon^{-1}$ in \cite{fk}) is less than about 10, most fragments are quasi-stable and will probably slowly evolve into Q-balls. However, if $r_{E}$ is 100 or more, then the initial oscillon-like fragments break up into Q-balls of positive and negative charge. Thus our assumption of evolution to Q-balls is consistent with the results of \cite{fk} if $r_{E} \lae 10$, but in most of our examples $r_{E}$ is between 10 and 100, where the evolution of the fragments is not clear. 
The non-linear dynamics of oscillon-like objects requires very high resolution simulations to evaluate their stability \cite{oscillons}. The spatial size of the simulations of \cite{fk} is sufficiently large to observe many condensate fragments, but this necessarily means that the resolution of a single fragment is not optimized; a high-resolution simulation of just one fragment would be the best way to clearly establish the value of $r_{E}$ for which the fragments are quasi-stable. The emission of oppositely charged Q-ball pairs might be interpreted as a way for the condensate fragments to lose excess energy, but this appears to be contradicted by the observation of quasi-stable fragments with $r_{E} \approx 10$ in the simulation of \cite{fk}, which suggests complex non-linear dynamics favouring spherically symmetric oscillon-like states. This is an interesting question for future research. 

         Is Q-ball decay in GMSB AD baryogenesis a testable model for the origin of gravitino dark matter? The model predicts a large messenger scale, which will be reflected in the spectrum of SUSY particles \cite{strumia}. Moreover, the 2 GeV gravitino mass predicted by the model is sufficiently large that there should be a significant gravity-mediated contribution to the spectrum of SUSY particle masses. The model also predicts that decay of the thermal relic NLSP density must underproduce gravitino dark matter when $m_{3/2}$ is assumed to be 2 GeV, in order that the contribution from thermal NLSP decay is less than that from Q-ball decay. These features may be testable at the LHC; in particular, identification of the 2 GeV gravity-mediated contribution to SUSY breaking masses would provide strong evidence\footnote{The phenomenological advantages of an O(1) GeV gravitino with respect to flavour problems and dark matter have been discussed in the context of "'sweet spot"  \cite{ss} and "Goldilocks" \cite{gold} supersymmetry.}.  The GMSB flat direction might also produce correlated baryon and dark matter isocurvature perturbations, if the GMSB potential in the angular direction is sufficiently flat and $|\Phi|$ has the right magnitude during inflation.

          In summary, we have shown that unstable Q-balls are a natural possibility in GMSB AD baryogenesis with a large messenger scale. Such unstable Q-balls are essential if AD baryogenesis in GMSB is to be consistent with astrophysical constraints from neutron star stability. The requirement of a large messenger scale is also consistent with the 2 GeV gravitino mass required if gravitino dark matter comes from Q-ball decay. This points to a striking consistency between the requirements for gravitino dark matter from Q-ball decay and successful Affleck-Dine baryogenesis in GMSB models.


\end{document}